\documentclass[longauth]{aa}
\usepackage{graphicx}
\usepackage{longtable}
\usepackage{multirow}
\usepackage{natbib}

\usepackage{txfonts} 

\bibpunct{(}{)}{;}{a}{}{,}
\begin{document}
\renewcommand{\topfraction}{0.85}
\renewcommand{\bottomfraction}{0.7}
\renewcommand{\textfraction}{0.15}
\renewcommand{\floatpagefraction}{0.66}

\title{
Search for Pulsed VHE Gamma-Ray Emission from Young Pulsars with H.E.S.S.
}

\def\g{\hbox{$\gamma$}}

\author{F. Aharonian\inst{1}
 \and A.G.~Akhperjanian \inst{2}
 \and A.R.~Bazer-Bachi \inst{3}
 \and M.~Beilicke \inst{4}
 \and W.~Benbow \inst{1}
 \and D.~Berge \inst{1} \thanks{now at CERN, Geneva, Switzerland}
 \and K.~Bernl\"ohr \inst{1,5}
 \and C.~Boisson \inst{6}
 \and O.~Bolz \inst{1}
 \and V.~Borrel \inst{3}
% ...
}

\institute{
Max-Planck-Institut f\"ur Kernphysik, P.O. Box 103980, D 69029
Heidelberg, Germany
\and
 Yerevan Physics Institute, 2 Alikhanian Brothers St., 375036 Yerevan,
Armenia
\and
Centre d'Etude Spatiale des Rayonnements, CNRS/UPS, 9 av. du Colonel Roche, BP
4346, F-31029 Toulouse Cedex 4, France
\and
Universit\"at Hamburg, Institut f\"ur Experimentalphysik, Luruper Chaussee
149, D 22761 Hamburg, Germany
\and
Institut f\"ur Physik, Humboldt-Universit\"at zu Berlin, Newtonstr. 15,
D 12489 Berlin, Germany
\and
LUTH, UMR 8102 du CNRS, Observatoire de Paris, Section de Meudon, F-92195 Meudon Cedex,
France
% ...
}

\date{Received ? / Accepted ?}

\offprints{Stefan Schlenker,\\
\email{stefan.schlenker@cern.ch}}

\abstract
{} 
%
{
We present the results of a search for pulsed very-high-energy (VHE)
\g-ray emission from young pulsars using data taken with the H.E.S.S.\
imaging atmospheric Cherenkov telescope system.
}
%
{
Data on eleven pulsars, selected according to their spin-down
luminosity relative to distance, are searched for \g-ray signals with
periodicity at the respective pulsar spin period. Special analysis
efforts were made to improve the sensitivity in the 100\,GeV \g-ray
energy domain in an attempt to reduce the gap between satellite and
ground-based \g-ray instruments.
}
%
{
No significant evidence for pulsed emission is found in any data set.
Differential upper limits on pulsed energy flux are determined for all
selected pulsars in the approximate \g-ray energy range between 100\,GeV and
$50$\,TeV, using different limit determination methods, testing a
wide range of possible pulsar light curves and energy spectra. 
}
%
{
The limits derived here imply that the magnetospheric VHE \g-ray
production efficiency in young pulsars is less than $10^{-4}$ of the
pulsar spin-down luminosity, requiring spectral turnovers for the
high-energy emission of four established \g-ray pulsars, and constrain
the inverse Compton radiation component predicted by several outer gap
models.
}




\keywords{Gamma-rays: observations -- Pulsars: individual: Crab pulsar, Vela pulsar, PSR\,B1259$-$63, PSR\,B1509$-$58, PSR\,B1706$-$44}

\authorrunning{F. Aharonian et al.}
\titlerunning{
H.E.S.S.\ Search for Pulsed VHE $\gamma$-rays from Young Pulsars}

\maketitle

\section{Introduction}

Rotating neutron stars are known to convert a significant part of
their rotational energy into radiation originating from within the
magnetosphere. This emission is observable as a periodic signal at the
neutron star rotation frequency (the \emph{pulsar} phenomenon). For
many of the known young and energetic pulsars, the emitted luminosity
peaks at X-ray or \g-ray energies \citep[see][ and references
therein]{egret_limits}, usually attributed to curvature radiation of
accelerated electrons in the strong magnetic fields pervading the
pulsar magnetosphere. The luminosity of the pulsed high-energy
emission was found to correlate significantly with the energy loss
rate of the pulsar, i.e.\ its spin-down power $\dot{E}$
\citep{gamma_energetics, Cheng}, which relates to the magnetic field
at the neutron star surface, $B_{\rm surf}$, approximately as
$\dot{E}\propto \Omega^4 B^2_{\rm surf}$, with $\Omega$ as the angular
frequency of the pulsar rotation.
For most of the pulsars with established \g-ray emission
\citep{egret_crabvela,1509_comptel}, there is evidence for a turnover
in the pulsed spectrum at a critical energy $E_{\rm c}$ in the sub-GeV
to 10\,GeV range. Note that no such turnover has been found up to
10\,GeV in the case of PSR\,B1706$-$44 \citep{egret_1706}.


Currently, two different scenarios for magnetospheric \g-ray emission
are discussed, placing the emission regions either near the magnetic
poles of the neutron star \citep[{\em polar cap} model,
][]{pc:0,pc:0a,pc:2,pc:3,pc:4} or near the null surface in the outer
magnetosphere of the pulsar \citep[{\em outer gap}
model,][]{og:1a,og:1b,ogChiang92,ogChiang94,ogRomani96,ogZhang97,ogCheng2000}.
Both models predict a cutoff in the curvature radiation spectrum at
\g-ray energies of the order of GeV up to several tens of
GeV. However, the cutoff is of a different physical nature in the two
models: while in the case of the polar cap model, the maximum energy
of curvature radiation photons escaping the magnetosphere is limited
by pair production in the polar magnetic field, the maximum energy of
the accelerated electrons determines the maximum photon energy in the
outer gap scenario. This results in different predicted light-curves,
spectral shapes, and cutoff energies $E_{\rm c}$ of the emitted
high-energy radiation in the two models. Note that the predictions of
outer gap models generally depend more strongly on model parameters,
such as the inclination between the magnetic and rotational axes and
the viewing angle to the observer, than is the case for polar cap
models. Additionally, in some outer gap model calculations, a spectral
component in the TeV range due to inverse Compton (IC) up-scattering
of soft ambient seed photons by the accelerated electrons is predicted
\citep[e.g.\ ][]{og_1509,og:2,og2d}. The boosted TeV photons, the flux
of which is determined by the seed photon density, are more likely to
escape due to the lower magnetic field strength in the outer
magnetosphere compared to the polar cap regions.

Bridging the space between the distinct polar cap and outer gap
models, particle acceleration within the slot gap \citep{slotgap}, a
narrower region between the last open magnetic field line and the
pair-plasma region of the polar cap, has been discussed in the context
of high-energy radiation. However, so far slot gap calculations have
been used primarily to model details of pulsed high-energy
light-curves \citep{sgDyksRudak2003,sgMuslimov2003} without
reproducing the corresponding energy spectra in the GeV range or
predicting spectra for TeV energies.


Pulsars with established \g-ray emission components have been
subject to intensive searches for pulsed very-high-energy (VHE,
energies above $\sim$\,$100$\,GeV) \g-ray emission by ground based
instruments. Up to now, no evidence for pulsed emission has been found
in these observations \citep{cangaroo_vela, durham, celeste_crab,
whipple_crab, hegra_crab:2}, and upper limits on the pulsed VHE \g-ray
flux have been derived under various assumptions on the characteristics of
the pulsed emission. However, the IC component predicted by outer gap
models has not yet been significantly constrained.


The superior sensitivity of the H.E.S.S.\ detector with respect to
previous instruments \citep[see e.g.][]{OngStatus}, combined with
its Southern Hemisphere location, puts the predicted pulsed IC
component from outer gap models within reach of testability,
motivating a new search for pulsed VHE emission from the known \g-ray
pulsars. A significant amount of H.E.S.S.\ data were accumulated in a
survey of the inner Galaxy \citep{scan2}, resulting in the detection
of numerous new sources of VHE \g-rays, many of which are apparently
associated with pulsars. Thus, in addition to the known \g-ray
pulsars, these data can be used to search for pulsed emission from
promising young, energetic pulsars.

The paper is structured as follows: in Section~\ref{sec_observations}
the selection criteria of the studied pulsar sample are explained,
Section~\ref{sec_analysis} introduces the H.E.S.S.\ instrument and
data analysis, in Section~\ref{sec_periodicity} the methods employed
in the search for variability and the flux limit determination are
explained, and finally, the results are presented and discussed in
Sections~\ref{sec_results} and \ref{sec_discussion}, respectively.

\section{Pulsar Candidate Selection}
\label{sec_observations}

The prime candidates for the search for VHE \g-ray emission are the
pulsars with established \g-ray emission at energies below
$\sim$\,$10$\,GeV which have been detected by CGRO instruments. Four
of them, PSR\,B0531$+$21 (Crab pulsar), PSR\,B0833$-$45 (Vela pulsar),
PSR\,B1706$-$44, and PSR\,B1509$-$58 were the subject of pointed
H.E.S.S.\ observations and many hours of observations were obtained
for each of them in the years 2002 to 2005. In addition, a significant
number of known pulsars were observed as part of the H.E.S.S.\ survey
of the Galactic Plane in 2004 and 2005, or lie within the field of
view of observations of other H.E.S.S.\ targets. The location of the
H.E.S.S.\ site in the southern hemisphere limits the declination of
observations to $<30\degr$.

Apart from the known \g-ray pulsars, other candidates for which
H.E.S.S.\ data were available were selected from the ATNF pulsar
catalogue \citep[see][ and references therein]{atnfcat} if their
spin-down flux $\dot{E}/D^2$ was greater than $10^{35}\,\rm
erg\,s^{-1}\,kpc^{-2}$.
Table~\ref{table_candidates} lists all candidates chosen along with
selected measured and derived characteristics collected from the
literature.

\begin{table*}[t]
\caption{\label{table_candidates}
The characteristics of the selected pulsars taken from
\citet{atnfcat}. Period, $P$, distance, $D$,
spin-down age, spin-down luminosity, $\dot{E}$, and the corresponding
value for $\dot{E}/D^2$, and calculated magnetic field strength at the
neutron star surface, $B_{\rm surf}$, and the light cylinder, $B_{\rm
LC}$, are listed. The last column shows the rank in $\dot{E}/D^2$
within the ATNF catalogue.}
\centering
\begin{tabular}{ccccccccccc}
\hline\hline
\multicolumn{2}{c}{Pulsar name} & $P$ & $D$ & Age & \multirow{2}*{$\displaystyle\log_{10}\left(\frac{\dot{E}}{\rm erg\,s^{-1}}\right)$} & \multirow{2}*{$\displaystyle\log_{10}\left(\frac{\dot{E}/{\rm erg\,s^{-1}}}{D^2/{\rm kpc^2}}\right)$} & $B_{\rm surf}$ & $B_{\rm LC}$ & Rank\\
\multicolumn{2}{c}{PSR} & $[\rm ms]$ & $[\rm kpc]$ & $[\rm kyears]$ & & & $[\rm 10^{11}\,G]$ & $[\rm 10^4\,G]$ & $\dot{E}/D^2$\\
\hline
B0531+21   & J0534+2200   & 33.1 & 2 & 1.24 & 38.7 &
38.1 & 37.8 & 98.0 & 1\\
B0833$-$45 & J0835$-$4510 & 89.3 & 0.29 & 11.3 & 36.8 &
37.9 & 33.8 & 4.45 & 2\\
B1706$-$44 & J1709$-$4429 & 102  & 1.8 & 17.5 & 36.5 &
36.0 & 31.2 & 2.72 & 6\\
B1509$-$58 & J1513$-$5908 & 151  & 4.4 & 1.55 & 37.3 &
36.0 & 154  & 4.22 & 7\\
& J1747$-$2958 & 98.8 & 2.5 & 25.5 & 36.4 &
35.6 & 24.9 & 2.42 & 13\\
B1259$-$63& J1302$-$6350 & 47.8 & 1.5 & 332  & 35.9 &
35.5 &  3.3 & 2.87 & 15\\
& J1811$-$1925 & 64.7 & 5  & 23.3 & 36.8 &
35.4 & 17.1 & 5.92 & 18\\
& J1524$-$5625 & 78.2 & 3.8 & 31.8 & 36.5 &
35.3 & 17.7 & 3.46 & 19\\
& J1420$-$6048 & 68.2 & 7.7 & 13   & 37.0 &
35.3 & 24.1 & 7.13 & 22\\
& J1826$-$1334 & 101  & 4.1 & 21.4 & 36.4 &
35.2 & 27.9 & 2.51 & 23\\
& J1801$-$2451 & 125  & 4.6 & 15.5 & 36.4 &
35.1 & 40.4 & 1.95 & 30\\
\hline
\end{tabular}
\end{table*}

Some peculiarities of individual pulsars are pointed out in the
following. PSR\,J1747$-$2958, situated in a bow-shock pulsar wind
nebula (PWN) is detected as a compact X-ray source \citep{fJ1747}. The
only binary pulsar in the present sample, PSR\,B1259$-$63 orbits a
Be-star companion. Its magnetic field is interestingly low
($\sim$\,$3\times$\,$10^{11}$\,G), suggesting a high cutoff energy in
polar cap models. It has been discovered as an emitter of transient
X-rays \citep{Cominsky:1} and VHE \g-rays \citep{1259}, however,
pulsed high-energy emission has not so far been
detected. PSR\,J1420$-$6048 is embedded in a PWN and identified with a
compact X-ray source \citep{fJ1420} within the Kookaburra complex,
from which VHE \g-ray emission was recently detected
\citep{Kookaburra}.
PSR\,J1811$-$1925, for which pulsed emission has been detected up to
hard X-rays \citep{fJ1811}, is located in the composite supernova
remnant G11.2$-$0.3. All of the selected pulsars are visible at radio
wavelengths with the exception of PSR\,J1811$-$1925
which is only seen in X-rays.

All available H.E.S.S.\ data were searched for the respective pulsar to be
in the field of view with a maximum angular distance of
$\theta=2$\degr\ between the pulsar position and the H.E.S.S.\
pointing direction.

\section{VHE Gamma-Ray Detection and Data Analysis}
\label{sec_analysis}

\subsection{The H.E.S.S.\ Detector}

The High Energy Stereoscopic System \citep[H.E.S.S.,][]{status}, an
array of imaging atmospheric Cherenkov telescopes located in Namibia
at $23\degr16\arcmin18\arcsec$\,S $16\degr30\arcmin00\arcsec$\,E and
1800\,m a.s.l., detects cosmic VHE \g-rays by imaging the Cherenkov
emission of their air showers in the atmosphere using optical
telescopes. Each telescope has a tessellated spherical mirror with
13\,m diameter and $107\,\rm m^2$ area \citep{optics1, optics2} and is
equipped with a camera of 960 $0.16\degr$-photomultiplier tubes
providing a total field of view of $5\degr$ in diameter
\citep{camera}. During the stereoscopic observations, an array trigger
requires the simultaneous detection of air showers by several
telescopes at the hardware level
\citep{trigger}, allowing a suppression of background events and the
assignment of GPS timestamps to each triggered air shower, without the
need of an offline synchronisation of the individual telescopes. After
digitisation, the event data are stored on disk by the data
acquisition system \citep{daq2003}.

Following the calibration of the recorded air shower data
\citep{calibration}, each telescope image was parametrised by its
centre of gravity and second moments \citep{Hillas} followed by the
stereoscopic reconstruction of the shower geometry providing an
angular resolution of $\sim0.1\degr$ for individual $\gamma$-rays.
The $\gamma$-ray energy was estimated from the image intensity and the
shower geometry with a typical resolution of $\sim15\%$. In order to
reject the vast background of cosmic-ray showers, $\gamma$-ray
candidates are selected using cuts on image shape scaled with their
expectation values obtained from Monte Carlo simulations. Details of
the analysis technique can be found in \citet{Crab}. The selection
cuts used for this analysis were optimised on simulations of a
$\gamma$-ray point source with 10\% of the flux of, and the same
spectral shape as, the VHE $\gamma$-ray emission from the Crab
Nebula. These cuts, in the following called {\em standard}, allow the
detection of sources with 1\% Crab Nebula flux at energies above
$\sim200\,$GeV with a significance of 5\,$\sigma$ within 25\,hours at
a zenith angle of 20$\degr$. Note that the threshold energy above
which \g-ray sources can be detected increases with the zenith angle
of observation.

\subsection{Low Energy Optimisation}
\label{sec_lowe_analysis}

\begin{figure}[t]
\centering
\resizebox{\hsize}{!}{\includegraphics{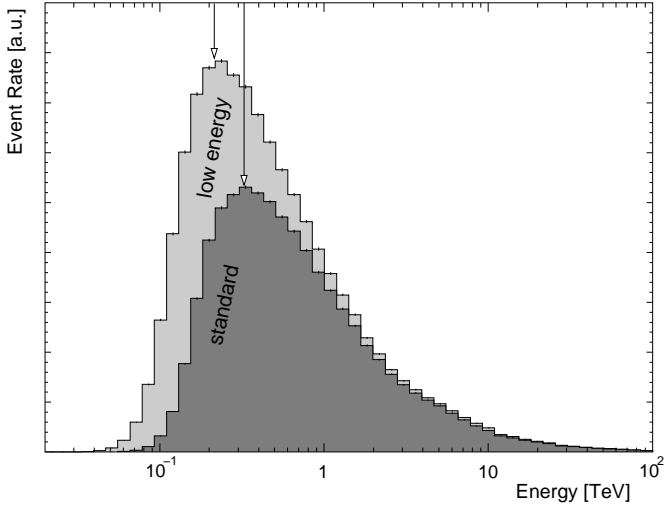}}
\caption{
Event rate of simulated \g-rays as a function of \g-ray energy for two
different sets of selection cuts. The dark grey histogram corresponds
to the standard and the light grey histogram to the low energy
cuts. The vertical arrows roughly represent the corresponding
threshold energy.}
\label{fig_thresholds}
\end{figure}

Since observational data indicate steep cut-offs in high-energy (GeV)
\g-rays, special {\em low-energy} cuts have been applied to reduce the
gap in observational coverage between satellite and ground based
\g-ray observations of young pulsars. Table~\ref{table_cuts} lists the
values of the two samples of selection cuts used in this analysis. See
\citet{Crab} for a detailed description of the individual cuts. The
low energy cuts were optimised on simulated \g-rays to yield the best
sensitivity for steep spectrum sources (${\rm d}N/{\rm d}E\propto
E^{-\Gamma}, \Gamma=5$) with energies below
500\,GeV. Figure~\ref{fig_thresholds} shows the rate of simulated
\g-ray events for a Crab-like source spectrum ($\Gamma=2.6$) passing the
standard (dark grey) and low energy (light grey) selection cuts as a
function of \g-ray energy for a typical zenith angle of 20\degr. The
peak of the distribution ${\rm d}N/{\rm d}E$ is commonly defined as
the threshold energy. The low energy cuts result in an energy
threshold lowered by a factor of 1.5 with respect to the standard
cuts. Although the number of selected events is higher by 50\%, the
background rejection efficiency of the low energy cuts is worse by
more than a factor of 3. Thus, the extension of the detectable \g-ray
energy range is achieved only at the expense of decreased sensitivity,
especially at higher energies.

\begin{table}[t]
\caption{\label{table_cuts}
Selection cuts used for the analysis of air shower events. The image
amplitude (Size) is given in photo-electrons (p.e.), $E_{\rm max}$
corresponds to the maximum reconstructed event energy, MRSW and MRSL
represent the mean scaled reduced width and length parameters, and
$\theta^2_{\rm cut}$ is the maximum squared angular distance with
respect to the target position. The tail cuts represent the image
cleaning thresholds applied. For details see \citet{Crab}.}
\centering
\resizebox{\hsize}{!}{
\begin{tabular}{cccccccccccc}
\hline\hline
Cuts & Size   & $E_{\rm max}$& MRSW & MRSL & $\theta^2_{\rm cut}$ & Tail cuts\\
     & [p.e.] & TeV          &      &      & $\rm [deg^2]$        & [p.e.]\\
\hline
Standard   & $>$80 & 100 & $[-2, 0.9]$  & $[-2, 2]$ & 0.0125 & 5/10\\
Low energy & $>$40 & 0.5 &$[-2, 0.75]$ & $[-2, 1.4]$ & 0.024  & 4/7 \\
\hline
\end{tabular}
}
\end{table}

\begin{figure}[t]
\centering
\resizebox{\hsize}{!}{\includegraphics{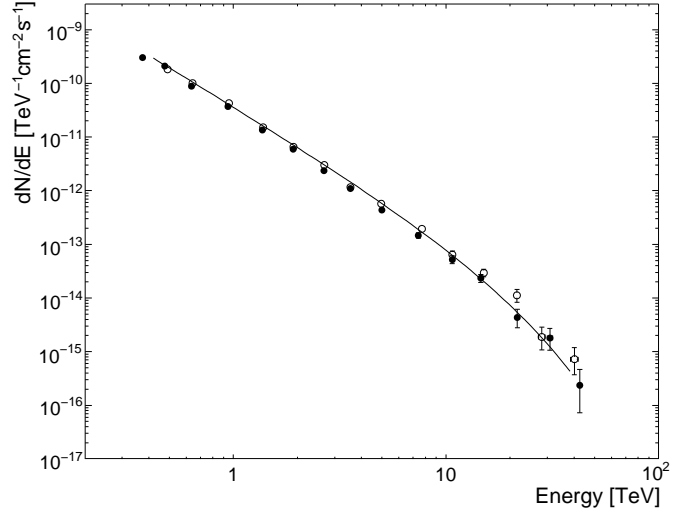}}
\caption{
Differential energy spectrum of unpulsed VHE \g-rays from the Crab
Nebula obtained with the different sets of selection cuts. The open
points correspond to the results of the standard and the full points
to the low energy cut analysis (but with $E_{\rm max}=100$\,TeV as for
the standard cuts). The solid line represents the spectrum
parametrisation taken from \citet{Crab}.}
\label{fig_crab_dcspec}
\end{figure}

In order to quantify possible systematic effects of the low energy
cuts on the energy spectrum determination, data from observations of
the Crab Nebula were used to reconstruct the differential energy
spectrum for both sets of cuts. In Figure~\ref{fig_crab_dcspec} the
open points show the spectrum of unpulsed \g-rays determined with the
standard cuts. The solid line indicates the best-fit power law with
exponential cutoff taken from \citet{Crab}, Table~5. The full points
result from the analysis with low energy cuts, showing minor
systematic differences in the absolute flux normalisation ($<$10\%).

\subsection{Point Source Analysis}
\label{seq_analysis}

Observation runs are selected according to standard quality selection
criteria, ensuring stable detector and atmospheric conditions as
described in \citet{Crab}. The data were analysed using the standard
method for the search for point sources without considering any time
variability. The number of events in a circular region around the
source position with radius $\theta_{\rm cut}$ is compared to the
number of counts in several background regions of the same size,
arranged on a circle around the centre of the field of view with
radius $\theta_{\rm off}$. For each run, the background normalisation
is determined by the number of background regions used. For data sets
comprised of several different offsets, as is the case for most of the
selected pulsars, the background normalisation for each run was
weighted by its relative acceptance for a given offset. This
acceptance was derived from data of H.E.S.S.\ observations without
significant \g-ray emission within the field of view \citep[see
also][]{background}.

\section{Periodicity Analysis}
\label{sec_periodicity}

In order to test for pulsed \g-ray emission at the pulsar position,
the timestamps of each recorded shower passing selection cuts were
transformed from the observer's frame into the pulsar frame and then
folded with the pulsar spin period. The resulting unbinned
distribution of pulsar phases corresponding to each shower event was
tested for variability. In cases where no significant periodicity was
found, flux limits were determined under various assumptions
concerning pulse shape and position.

\subsection{Timing Corrections}

The time of arrival (TOA) for each shower event recorded by the
telescope array is determined with a central trigger GPS clock
\citep{trigger}, achieving an accuracy of better than $2\,\mu $s. Each
TOA $t_{\rm TOA}$ was transformed to the pulsar frame using the
following corrections:
\begin{equation}
t = t_{\rm TOA} + \Delta t_{\rm clock}
 + \Delta t_{\rm SSB} + \Delta t_{\rm Shapiro} + \Delta t_{\rm Einstein}
 + \Delta t_{\rm binary}.
\end{equation}
Here, $\Delta t_{\rm clock}$ represents the transformation from GPS
time to International Atomic Time, $\Delta t_{\rm SSB}$ the classical
transformation to the solar system barycentre using the solar system
ephemerides DE200 from the Jet Propulsion Laboratory \citep{de200} and
$\Delta t_{\rm Shapiro}$ and $\Delta t_{\rm Einstein}$ are general
relativistic corrections for the gravitational influence of the sun
\citep{shapirodelay} and the redshift induced by all major bodies in
the solar system \citep{einsteindelay}, respectively. The correction
term $\Delta t_{\rm binary}$ applies only for pulsars in binary
systems, i.e.\ PSR\,B1259$-$63, and corrects for the pulsar motion
in its binary orbit according to \citet{btbinary}. The correction
software used for this analysis was cross-checked with the commonly
used pulsar timing utility TEMPO \citep{tempo}.
Furthermore, the timing capability of H.E.S.S.\ was demonstrated by
measuring the \emph{optical} light-curve of the Crab pulsar with high
precision \citep{optical}.

\subsection{Tests for Pulsed Emission}
\label{test}

The corrected TOA was subsequently folded with the timing parameters
of the target pulsar (pulsar ephemeris) obtained from radio or
X-ray observations. For each event TOA $t$, the pulsar phase $\phi$
was determined according to
\begin{equation}\label{eq_phase}
\phi(t) = N_{\rm r}(t) - \lfloor N_{\rm r}(t) \rfloor
\end{equation}
where $\lfloor\ldots\rfloor$ denotes the floor function, and with
$N_{\rm r}$ as the number of pulsar rotations
\begin{equation}\label{eq_phase_expansion}
N_{\rm r}(t) = N_{\rm r}(t_0) + \nu\;\!|_{t_0}(t-t_0) +
\frac{1}{2}\dot{\nu}\;\!|_{t_0}(t-t_0)^2 +
\frac{1}{6}\ddot{\nu}\;\!|_{t_0}(t-t_0)^3
\end{equation}
where the parameters $t_0$, $N_{\rm r}(t_0)$, $\nu|_{t_0}$,
$\dot{\nu}|_{t_0}$, and $\ddot{\nu}|_{t_0}$ are the reference time,
the corresponding number of rotations, frequency, and first and second
frequency derivative, respectively, which altogether represent the
timing parameters of the pulsar ephemeris. Table~\ref{table_timing}
lists the references for the ephemeris used for each pulsar along with
the respective reference times $t_{0}$ depending on the number of
different parameter sets.

\begin{table}[h]
\caption{\label{table_timing}
Pulsar ephemerides. The first and last reference times ($t_{0}$) of
the timing solutions used in this analysis are given. For a complete
list of the solutions and the corresponding pulsar parameters see the
references: 1: \citet{jbcrab}, 2: \citet{atnfarchive}, 3:
\citet{J1747}.
}
\centering
\begin{tabular}{cccc}
\hline\hline
Pulsar & $t_0$ [MJD] & References\\
\hline
B0531+21     & $52929,\ldots,53446$      & (1)   \\
B0833$-$45   & $53123,\ldots,53469$      & (2) \\
B1259$-$63   & $53108,\ldots,53386$      & (2) \\
J1420$-$6048 & 53156, 53493              & (2) \\
B1509$-$58   & $53111,\ldots,53385$      & (2) \\
J1524$-$5625 & 53650                     & (2) \\
B1706$-$44   & 52773, 52915              & (2)   \\
J1826$-$1334 & 53130, 53449              & (2) \\
\hline
J1747$-$2958 & 52613$^\star$             & (2)   \\
J1801$-$2451 & 52503$^\star$             & (3)   \\
\hline
\multicolumn{3}{l}{
\begin{minipage}[t]{6.cm} 
$^\star$Valid timing information was not available for all data within the H.E.S.S.\ data set.\\
\end{minipage}
}\\
\end{tabular}
\end{table}

For the X-ray pulsar PSR\,J1811$-$1925
and the radio pulsars PSR\,J1801$-$2451 and PSR\,J1747$-$2958 there
were no contemporaneous observations available from which the pulsar
timing parameters could have been deduced. For these pulsars, the most
recently published timing solution was used, assuming that the pulsar phase
determined with the known parameters is accurate enough to reveal the
periodicity of a potential signal. In particular, this assumes that no
glitches or other major frequency changes occurred within the time
range between the H.E.S.S.\ observations and the measurement of the
timing parameters.

The unbinned phase distribution of events was tested for significant
periodicity, i.e.\ for a significant deviation from a flat
distribution, using several statistical tests. The Pearson
$\chi^2$-test tests for a constant distribution of phases in a number
of phase bins, which was set to 20 in order to give acceptable
coverage for any of the considered data sets. The $Z^2_m$-test
\citep{Zmtest} checks for a combination of fundamental sine and cosine
harmonics up to the $m$-th order and is independent of any
binning. Here, the $Z^2_1$- and $Z^2_2$-tests for single and double
peak profiles are employed. Additionally the $H$-test \citep{Htest}, a
more powerful generalisation of $Z^2_m$, was performed to increase the
sensitivity to unknown pulse profiles. Finally, a rather different
approach which checks for maximum deviations from the uniform phase
distribution, the Kuiper-test \citep{Ktest}, was used.

In order to test for spurious signals induced by possible systematic
effects in the H.E.S.S.\ event data, the tests were also applied to
the phase series of background events, i.e.\ events from sky regions
used in the background determination for the unpulsed \g-ray excess
(cf.\ Sec.~\ref{seq_analysis}).

\begin{figure}[t]
\centering
\resizebox{\hsize}{!}{\includegraphics{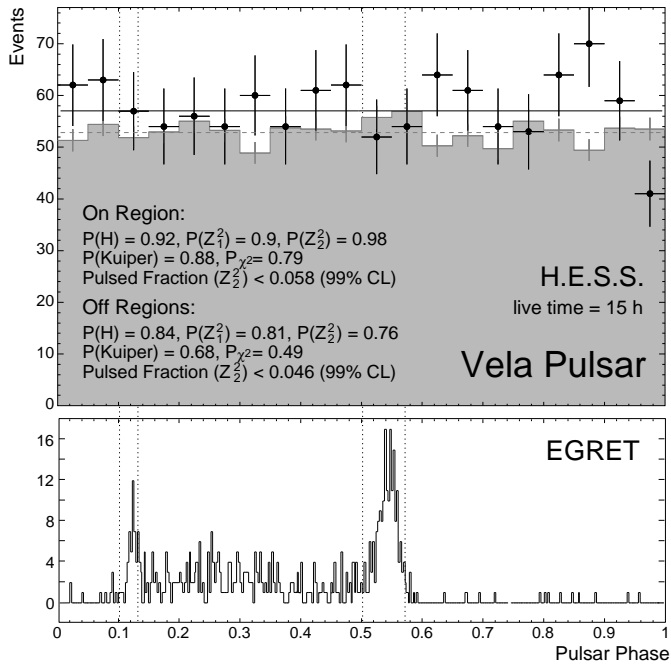}}
\caption{
{\em Upper plot}: Distribution of event phases for the Vela pulsar
(PSR\,B0833$-$45). The points represent the events in the on-region
at the pulsar position and the histogram the normalised off-region
events. The solid and dashed lines represent the $\chi^2$ fits to a
constant for the on- and off-events, respectively. For both regions
the probabilities for being consistent with a uniform distribution
according to the statistical tests on pulsations (see text) are listed
along with the pulsed fraction limit as introduced in
Sec.~\ref{sec_limits}. {\em Lower plot}: Phase distribution for
\g-rays with energies between 2 and 10\,GeV as measured by EGRET
\citep{egret_crabvela}. The dotted vertical lines denote the edges of the
on-pulse regions.}
\label{fig_phaso}
\end{figure}

As an example, the distribution of event phases from observations of
the Vela Pulsar (PSR\,B0833$-$45) is shown in Fig.~\ref{fig_phaso},
obtained using the standard cuts. The difference between on and off
results from the known \g-ray excess from HESS\,J0835$-$456 at the
position of the pulsar (see Table~\ref{table_dc_results}). No
significant deviation from uniformity was found with any of the
statistical tests for pulsations.

\subsection{Flux Limit Determination}
\label{sec_limits}

Several methods were applied to obtain limits on the \g-ray flux from
the selected pulsars. They differ in the assumptions made concerning
the characteristics of the pulsed emission, in particular concerning
the form of the light-curve and the phase position.

In the most conservative approach, without any assumptions on the
temporal structure or spectrum of a possible signal, the energy flux
$E^2{\rm d}N/{\rm d}E$ is calculated using the standard analysis as
described in Sec.~\ref{seq_analysis}. This method considers all events
in the direction of the pulsar, i.e.\ in the on-region, ignoring any
possible variations and is referred to as the {\em unpulsed}
method. The flux is calculated event-wise for the on- and off-
(background) regions within the same field of view separately,
accumulated in bins of energy, and subsequently subtracted to yield
the excess flux \citep{PKS2155}. Note that due to the finite width of
the energy bins, the reconstructed flux is weakly dependent on the
\g-ray energy spectrum for which the detector efficiency was
calculated. Finally, the flux obtained in this way is converted into a
99\% confidence level upper limit using the method of
\citet{Feldman:1}. As some of the pulsars in this sample show
significant unpulsed \g-ray emission, the flux limits derived can be
considerably higher than the nominal H.E.S.S.\ \g-ray flux sensitivity
since the unpulsed \g-ray emission is an additional background to the
pulsed emission.

The second method takes advantage of the pulse position and shape as
measured in other energy domains, assuming similar characteristics for
the hypothetical VHE \g-ray emission. In particular for the known
\g-ray pulsars this approach seems reasonable, since the \g-ray
emission mechanism is expected to be similar at VHE \g-ray energies,
especially for photons originating from curvature radiation. For each
pulsar, a known pulse profile was chosen according to the energy band
of the emission, taking the highest photon energies. For these pulse
profiles, an on- and an off-phase domain were assigned ({\em
on-off-pulse} method), representing the pulsar phases for high and low
(or no) emission, respectively. Table~\ref{table_phaseregions} lists
the used on-phase intervals and their corresponding energy bands for
all pulsars. The phase intervals are given relative to the reference
phase which is coincident with the highest peak of the respective
radio profile. For example for the Vela pulsar, the on- and off-phase
domains were chosen based on the light-curve measured using EGRET at
GeV energies \citep{egret_crabvela}, as indicated by the dotted lines
in Fig.~\ref{fig_phaso}. Note that any possible contribution from
inter-pulse domains, the phase range between the on-pulse domains, to
the pulsed signal are neglected in this method. For the pulsars
without established \g-ray emission, the selected phase region always
covers 20\% of the profile around the reference phase. The VHE \g-ray
flux was then calculated as for the unpulsed method, but using the
off-phase events in the direction of the pulsar for background
estimation, instead of the events in the off-regions within the field
of view. If there was no significant unpulsed signal visible within a
data set and no known VHE \g-ray source coincident with the target
pulsar, i.e.\ for PSR\,J1524$-$5625, PSR\,B1706$-$44, and
J1801$-$2451, the off-event statistics were increased by including
those events which were in the off-regions of the field of view and
additionally had phases falling in the assigned off-phase domain. In
these cases the background normalisation was adjusted accordingly.
\begin{table}[h]
\caption{\label{table_phaseregions}
Pulse phase regions used in the on-off-pulse method. Given are the
phase intervals considered as {\em on phase} together with the energy
band and the corresponding reference. References: 1:
\citet{egret_crabvela}, 2: \citet{parkes3}, 3: \citet{1509_comptel},
4: \citet{egret_1706}, 5: \citet{J1747}, 7: \citet{J1811}.
}
\centering
\begin{tabular}{cccc}
\hline\hline
Pulsar & On pulse phases & Energy & Reference\\
\hline
B0531+21   & $[-0.06,0.04],[0.32,0.43]$ & \g-ray& (1) \\
B0833$-$45 & $[0.1,0.14],[0.5,0.57]$    & \g-ray& (1) \\
B1259$-$63 & $[-0.1,0.1]$               & Radio & (2) \\
J1420$-$6048 & $[-0.1,0.1]$               & Radio & (2) \\
B1509$-$58 & $[0.15,0.65]$              & \g-ray& (3) \\
J1524$-$5625 & $[-0.1,0.1]$               & Radio & (2) \\
B1706$-$44 & $[0.24,0.5]$               & \g-ray& (4) \\
J1826$-$1334 & $[-0.1,0.1]$               & Radio & (2) \\
\hline
\end{tabular}
\end{table}
The on-off-pulse method
is less reasonable for those pulsars for which pulsed emission is
known only at low energies, especially at radio wavelengths, since the
\g-ray production mechanism may be completely unrelated to that of the low
energy emission.

Therefore, a third approach was also applied, that does not rely on
knowledge of the {\em phase position} of the pulse, but only assumes a
certain {\em pulse shape}. Here, a combination of first and second
order harmonics was chosen, similar to the $Z^2_2$-test. A 99\%
confidence level upper limit on the fraction, $p$\,$\ll$\,$1$, of
pulsed emission present in the series of event phases for events with
shower directions from within the on-region around the pulsar position
was calculated for a given energy bin according to
\citet{Brazier}. The resulting event fraction was converted into a
flux using the effective collection area of the detector and assuming
a spectral distribution in energy following a power law with photon
index $\Gamma$\,$=$\,$2.7$, roughly matching the shape of the cosmic
ray background spectrum. More details on the flux calculations can be
found in \citep{Crab}. This method, in the following referred to as
the {\em pulsed fraction} method, is less constraining than the
on-off-pulse method, since no background subtraction can be performed
and thus the post-cut cosmic ray background is treated as an unpulsed
component in addition to any unpulsed VHE \g-ray excess.

Altogether, the pulsed fraction method was considered to be the most
robust approach since the structure of the VHE \g-ray light curve
remains unknown a priori. Nevertheless, the on-off-pulse method was
still applied to ensure comparability with results from prior searches
for pulsed VHE \g-ray emission where very similar techniques have been
used. For the three pulsars for which valid timing parameters were not
available for the H.E.S.S.\ data, PSR\,J1747$-$2958, PSR\,J1801$-$2451
and PSR\,J1811$-$1925, flux limits have been derived only using the
unpulsed method.

\section{Results}
\label{sec_results}

\subsection{Point Source Analysis}

Table~\ref{table_dc_results} shows the main characteristics of the
data sets for each pulsar along with the results of the point source
analysis for unpulsed emission. Additionally, any H.E.S.S.\ VHE
\g-ray source close to the corresponding pulsar position is listed
whenever a significant \g-ray excess was detected which can be
attributed to this known source. Note that the excess given may differ
significantly from that given in the respective reference, since in
all cases, with the exception of the Crab nebula, the VHE emission
region is both extended and offset from the pulsar position.

\begin{table*}[t]
\caption{\label{table_dc_results}
Results of the analysis of unpulsed \g-ray emission for each pulsar
data set using the standard cuts and assuming a point source at the
pulsar position. Dead-time corrected observation time, $t_{\rm obs}$,
mean offset of the pulsar direction with respect to the array pointing
direction, $\langle\theta_{\rm off}\rangle$, approximate mean energy
threshold, $\langle E_{\rm thresh}\rangle$, number of excess counts,
$N_{\rm excess}$, and corresponding significance, $S$, are given. In
addition to the mean threshold for the standard cuts, the
corresponding threshold for the low energy cuts is given in
parentheses if applicable. If the pulsar is in the vicinity (angular distance $<1$\degr) of,
or associated with, a known VHE \g-ray source, the source name is
given along with the corresponding reference. Note that the given
numbers can significantly differ from the published results for the
known sources which are located at different positions and might be
extended. References: A: \citet{Crab}, B: \citet{VelaX}, C:
\citet{1259}, D: \citet{Kookaburra}, E: \citet{MSH}, F: \citet{1706},
G: \citet{scan2}, H: \citet{1825}.}
\centering
\begin{tabular}{cccccccccccc}
\hline\hline
Pulsar & $t_{\rm obs}$ & $\langle\theta_{\rm off}\rangle$ & $\langle
E_{\rm thresh}\rangle$ & $N_{\rm excess}$ & $S$ & H.E.S.S. &
Reference\\ PSR & [h] & [\degr] & [GeV] & & $[\sigma]$ & catalogue
source &\\
\hline
B0531+21   & 25.4 & 0.6 & 510 (310) & $9673\pm59$ & 164.8 & 
Crab Nebula & (A)\\
B0833$-$45 & 16.3 & 0.6 & 250 (170) & $97\pm40$   & 2.81  & 
HESS\,J0835$-$456 & (B)\\
B1259$-$63 & 55.8 & 0.6 & 380 (280) & $1081\pm65$ & 16.7  & 
HESS\,J1303$-$638 & (C)\\
J1420$-$6048 & 13.0 & 0.8 & 350 (240) & $338\pm 28$ & 12.07 & 
HESS\,J1420$-$607 & (D)\\
B1509$-$58 & 35.8 & 0.6 & 310 (210) & $1853\pm 58$& 32.22 & 
HESS\,J1514$-$591 & (E)\\
J1524$-$5625 & 15.5 & 1.0 & 330 (220) & $-8\pm28$   & $-0.30$ & --\\ 
B1706$-$44 & 16.6$^\star$ & 0.5 & 255 &  $1\pm24$  & 0.3 & --
& (F)\\ 
J1747$-$2958 & 103.3& 1.1 & 200 (140) & 291$\pm$82  & 3.62 & 
HESS\,J1745$-$303 & (G)\\
J1801$-$2451 & 9.8  & 1.2 & 180 (130) & $-19\pm 28$ & $-0.71$ & --\\ 
J1811$-$1925 & 14.2 & 1.1 & 210 (150) & $116\pm 31$ & 3.8$^\star$$\!^\star$ &
-- \\ 
J1826$-$1334 & 62.8 & 1.1 & 210 (140) & $683\pm63$  & 11.2 & 
HESS\,J1825$-$137 & (H)\\
\hline
\hline
\multicolumn{7}{l}{
\begin{minipage}[t]{15cm} 
$^\star$Comprises 14.4\,h of 2-telescope data 
and 2.2\,h of data from the full 4-telescope array.\\
$^\star$$\!^\star$See text.
\end{minipage}
}\\

\end{tabular}
\end{table*}

Three of the four pulsars not coincident with known VHE \g-ray sources
-- PSR\,J1524$-$5625, PSR\,B1706$-$44, and PSR\,J1801$-$2451 -- show
no significant signal, but for PSR\,J1811$-$1925
an excess with 3.8\,$\sigma$
significance is seen.
After taking into account all four trials for these target positions,
the significance is reduced to 3.5\,$\sigma$.

Since the H.E.S.S.\ energy threshold increases with the zenith angle
of observation, the low energy selection cuts introduced in
Sec.~\ref{sec_lowe_analysis} effectively reject all data taken at
zenith angles higher than 50\degr\ since the effective collection area
of the instrument vanishes below 500\,GeV \g-ray energy for these
elevations. Therefore the effective exposure time for the low energy
analysis is reduced for some data sets compared to that of the
standard cuts.

For PSR\,B1706$-$44, only archival data from the early commissioning
phase of H.E.S.S.\ from observations with two telescopes and without
an array level trigger are available. Therefore, event timing was
determined from the GPS timestamps of the individual
cameras. Additionally, the low energy analysis was not performed.

In the special case of PSR\,B1259$-$63, the data set used here
differs from that used in \citet{1259}.  Newer data from observations
in 2005 are included and data taken before MJD\,53135 are excluded.
Before this date reliable radio ephemerides could not be provided as
the radio pulsar was eclipsed by its binary companion. Additionally,
the unpulsed \g-ray emission, presumably arising from the interaction
of the pulsar wind and the stellar outflow, would result in a strong
background for any pulsed emission component.

\subsection{Search for Pulsations}

\begin{figure}[t]
\centering
\resizebox{\hsize}{!}{\includegraphics{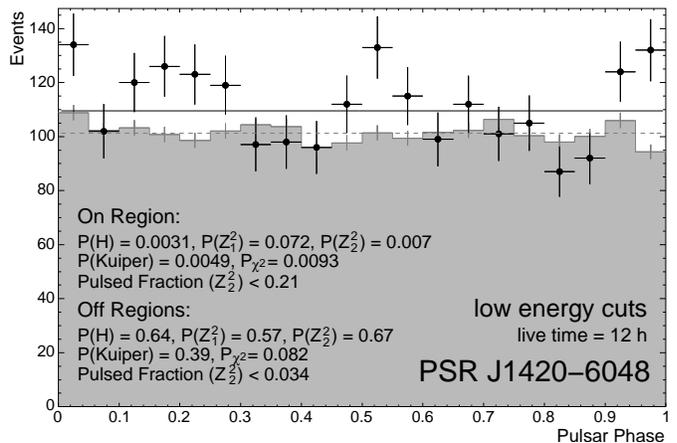}}
\caption{
Distribution of event phases for PSR\,J1420$-$6048 using the low
energy cut analysis for on- and off-regions (see also
Fig.~\ref{fig_phaso}). Note that there is significant unpulsed
emission at the pulsar position resulting in a difference between the
mean number of events for the on and off-regions shown by the solid
and dashed lines respectively.}
\label{fig_phaso_1420}
\end{figure}

Table \ref{table_stat_tests} shows the results of the statistical
tests applied to the phase distributions of events in the on-region
around each pulsar position. The probabilities for the hypothesis of a
flat phase distribution, i.e.\ the probability that no pulsed signal
exists, are given for the two sets of selection cuts applied. It can
be seen that these probabilities lie well above the one per cent level
with a single exception. For the low energy analysis of
PSR\,J1420$-$6048, the test probabilities seem significantly low,
especially for the more powerful H- and Kuiper-tests. No such effect
is seen for the respective probability of the background distributions
of any data set, which are all above 1\%. In
Figure~\ref{fig_phaso_1420}, the corresponding phase distribution is
shown, again for both on and off-events. The off-event distribution is
much more compatible with a flat distribution, with probabilities well
above 10\%. The probability $P_{\rm H}|_{\rm PSR\,J1420-6048} =
3.1\times10^{-3}$ of the H-test for the on-events corresponds to
$3.0\,\sigma$. However, this probability has to be corrected for the
number of trials for this test. Considering 19 trials for all data
sets analysed with the low-energy (9 trials) and standard (10 trials)
selection cuts respectively, the corresponding significance for the
applied tests lies between $2.0\,\sigma$ (H-test) and $0.4\,\sigma$
($Z^2_1$-test) with an average of $1.5\,\sigma$. Note that the test
results for the pulsars PSR\,J1747$-$2958 and PSR\,J1801$-$2451 for
which no valid timing information was available do not allow firm
conclusions about the presence of periodicity within the H.E.S.S.\
data with respect to the intrinsic pulsar rotation period. For
PSR\,J1811$-$1925, the tests were not performed since there was no
valid timing information available for any period within several years
of the H.E.S.S.\ observation time.

\begin{table*}[t]
\caption{\label{table_stat_tests}
Results of statistical tests for pulsed emission. Number of events,
$N_{\rm on}$, $\chi^2$ probability for a fit of the phase distribution
to a constant, and probabilities for the event series being compatible
with a flat phase distribution for the $H$-, $Z^2_1$-, $Z^2_2$-, and
Kuiper-test ($P_K$) are shown. All quantities are given for both the
low energy and standard cut analyses (see text).}
\centering
\begin{tabular}{c|cccccc|cccccc}
\hline\hline
Pulsar & \multicolumn{6}{c|}{Low energy cuts} & \multicolumn{6}{c}{Standard cuts}\\
PSR & $N_{\rm on}$ & $P_{\chi^2}$ & $P_H$ & $P_{Z^2_1}$ & $P_{Z^2_2}$ & $P_K$
    & $N_{\rm on}$ & $P_{\chi^2}$ & $P_H$ & $P_{Z^2_1}$ & $P_{Z^2_2}$ & $P_K$\\
\hline
B0531+21   & 8095  & 0.99  & 0.84 & 0.81 & 0.94 & 0.97  & 10622 & 0.51 & 0.56 & 0.63 & 0.25 & 0.66\\
B0833$-$45 & 7480  & 0.52  & 0.87 & 0.85 & 0.37 & 0.82  & 1156  & 0.79 & 0.92 &  0.90 & 0.98 & 0.88\\
B1259$-$63 & 16176 & 0.78  & 0.92 & 0.90 & 0.65  & 0.71  & 4535  & 0.29 & 0.25 & 0.18 & 0.46 & 0.23\\
J1420$-$6048 & 2228  & 0.0093& 0.0031&0.072& 0.007& 0.0049& 968   & 0.67 & 0.60 & 0.53 & 0.53 & 0.62\\
B1509$-$58 & 12481 & 0.37  & 0.81 & 0.77 & 0.70  & 0.89  & 4308  & 0.048& 0.055& 0.027& 0.11 & 0.04\\
J1524$-$5625 & 2498  & 0.78  & 0.43 & 0.35 & 0.43  & 0.39  & 745   & 0.87 & 0.79 & 0.75 & 0.96 & 0.97\\
B1706$-$44$^\star$ & --   & --     & --   & --   & --   & --    & 391   & 0.02 & 0.82 & 0.78 & 0.85 & --  \\
J1826$-$1334 & 14497 & 0.71  & 0.46 & 0.38 & 0.57  & 0.62  & 4016  & 0.46 & 0.42 & 0.34 & 0.34 & 0.42\\
\hline
J1747$-$2958$^\star$$\!^\star$ & 23482 & 0.65  & 0.98 & 0.97 & 0.82  & 0.95  & 6340  & 0.62 & 0.62 & 0.92 & 0.99 & 0.96\\
J1801$-$2451$^\star$$\!^\star$ & 3230  & 0.035 & 0.22 & 0.15 & 0.42  & 0.21  & 723   & 0.50 & 0.15 & 0.094& 0.24 & 0.32\\
\hline
\multicolumn{13}{l}{
\begin{minipage}[t]{15cm} 
$^\star$Subject to analysis limitations (archival data, see text).\\
$^\star$$\!^\star$Valid pulsar timing information was not available for all data within the H.E.S.S.\ data set.
\end{minipage}}
\\
\end{tabular}
\end{table*}

\begin{figure}[t]
\centering
\resizebox{0.95\hsize}{!}{\includegraphics{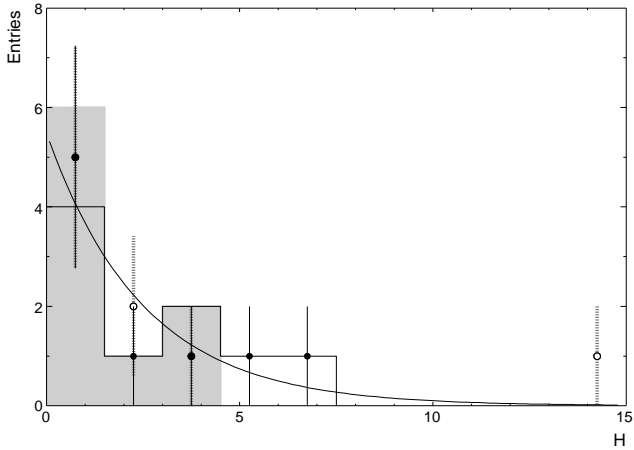}}
\caption{
Ensemble distributions of the H-test statistic for the selected
pulsars of Table~\ref{table_stat_tests} and their corresponding
background control samples. The results for the pulsar direction are
shown as open and closed circles for the low energy and standard cut
analysis, respectively. The distributions for the off-regions are
displayed as grey filled and outlined histograms, respectively. The
solid curve shows the expected distribution if no pulsed signals are
present.}
\label{fig_hdist}
\end{figure}

As a cross-check for systematic effects possibly biasing the
statistical tests, the ensemble distribution of the test statistic was
checked against the hypothesis that no pulsed signal exists in any of
the data sets. Figure~\ref{fig_hdist} shows the distribution of the
test statistic $H$ of the H-test for the on- and off-regions and for
the different sets of cuts. Also shown is the expected distribution
$N_{\rm H}(H) = N_0 \exp(-\lambda H)$ for the case when no pulsed
signal is present in any data set, with $N_{0}$ depending on the
number of tests and $\lambda=0.4$. In order to quantitatively evaluate
the match between data and expectation, each of the distributions for
on- and off-events and for the two sets of cuts was fitted by the
respective expected distribution with a log-likelihood method, leaving
$N_{0}$ and $\lambda$ as free parameters, and the probability for
being compatible with the expected parameters was calculated. No
significant deviation is found for any of the four distributions with
all probabilities above 20\%.

\subsection{Flux Upper Limits for Pulsed Emission}

\begin{figure}[ht]
\centering
\resizebox{\hsize}{!}{\includegraphics{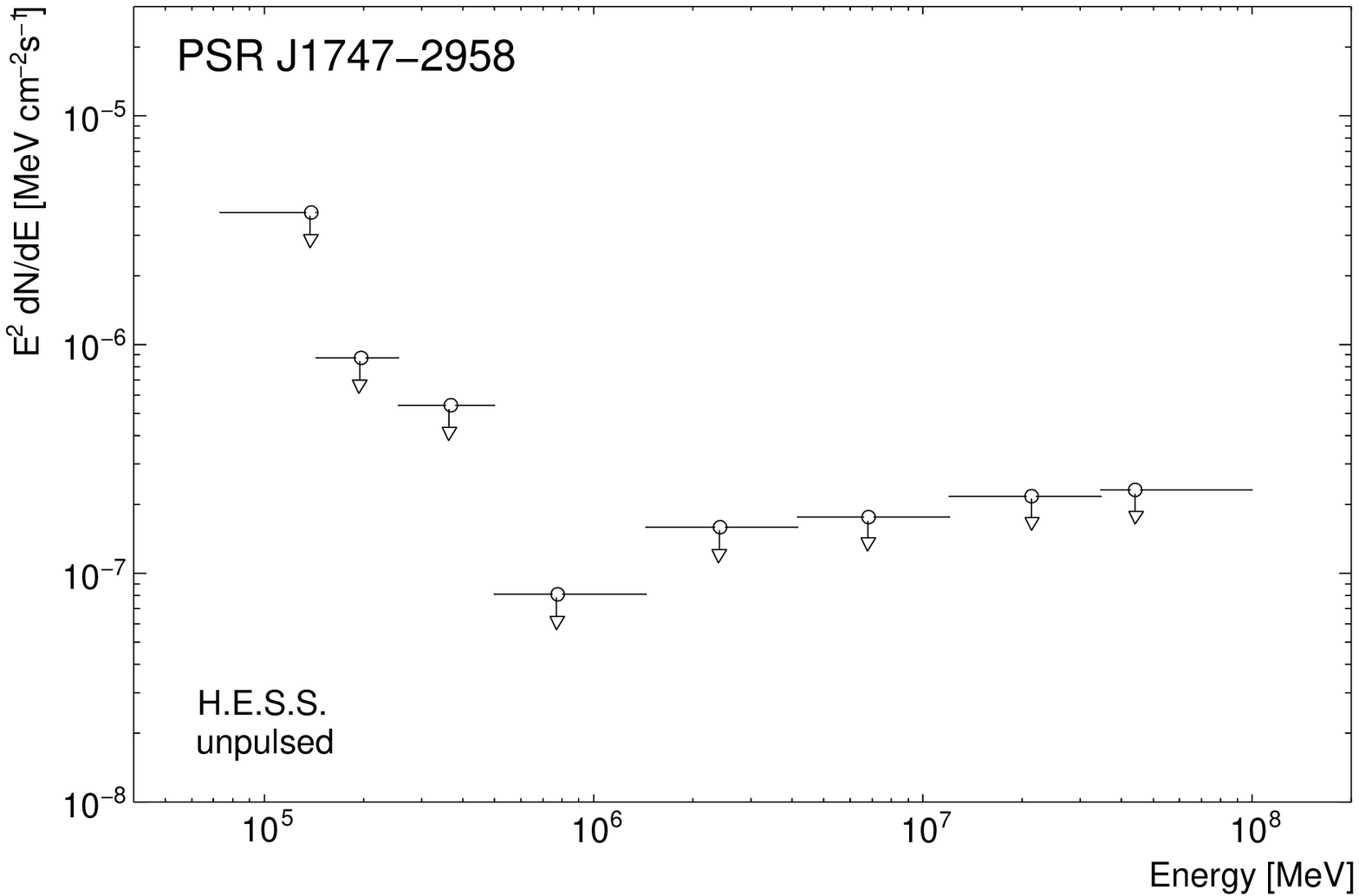}}
\resizebox{\hsize}{!}{\includegraphics{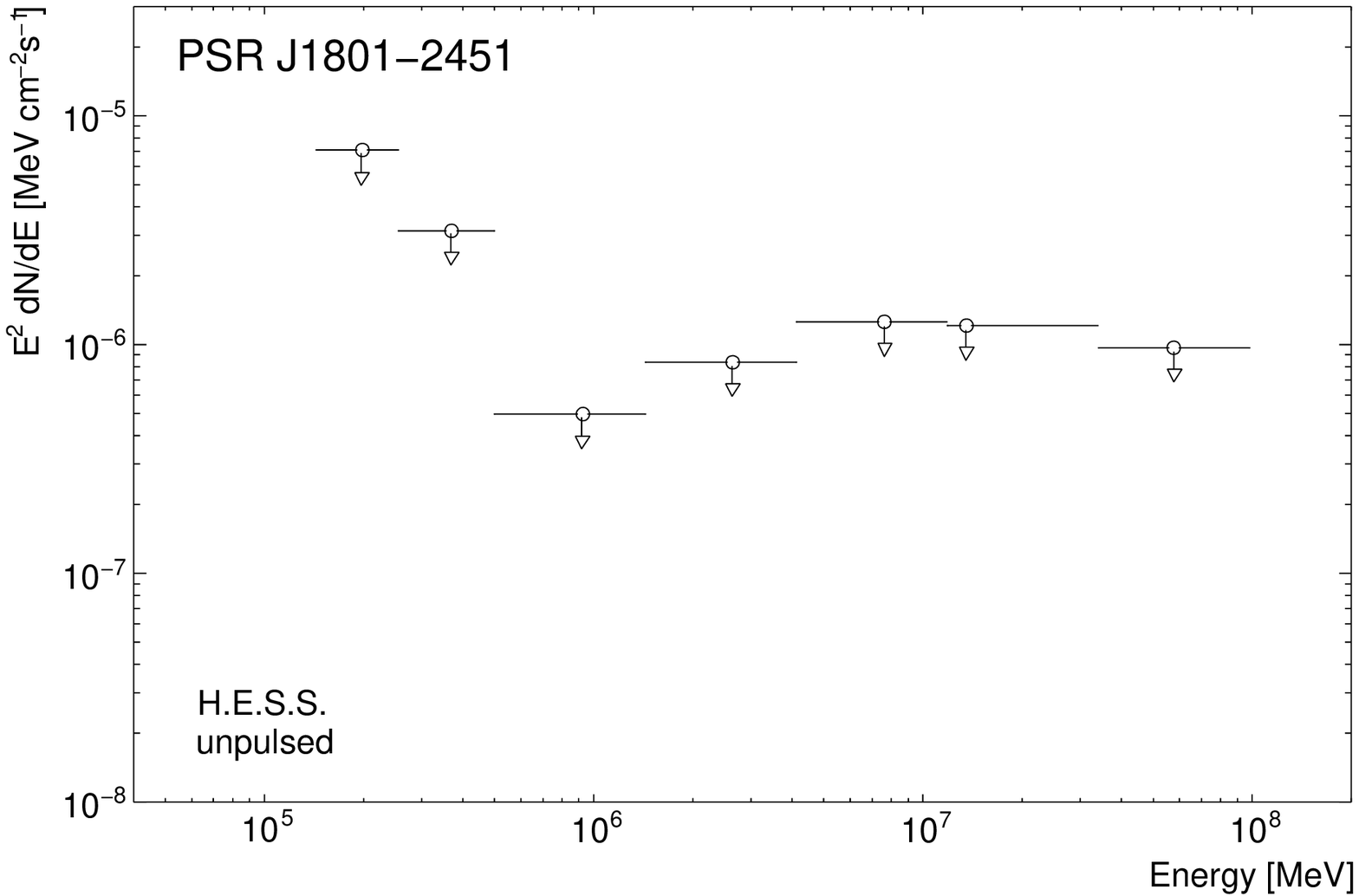}}
\resizebox{\hsize}{!}{\includegraphics{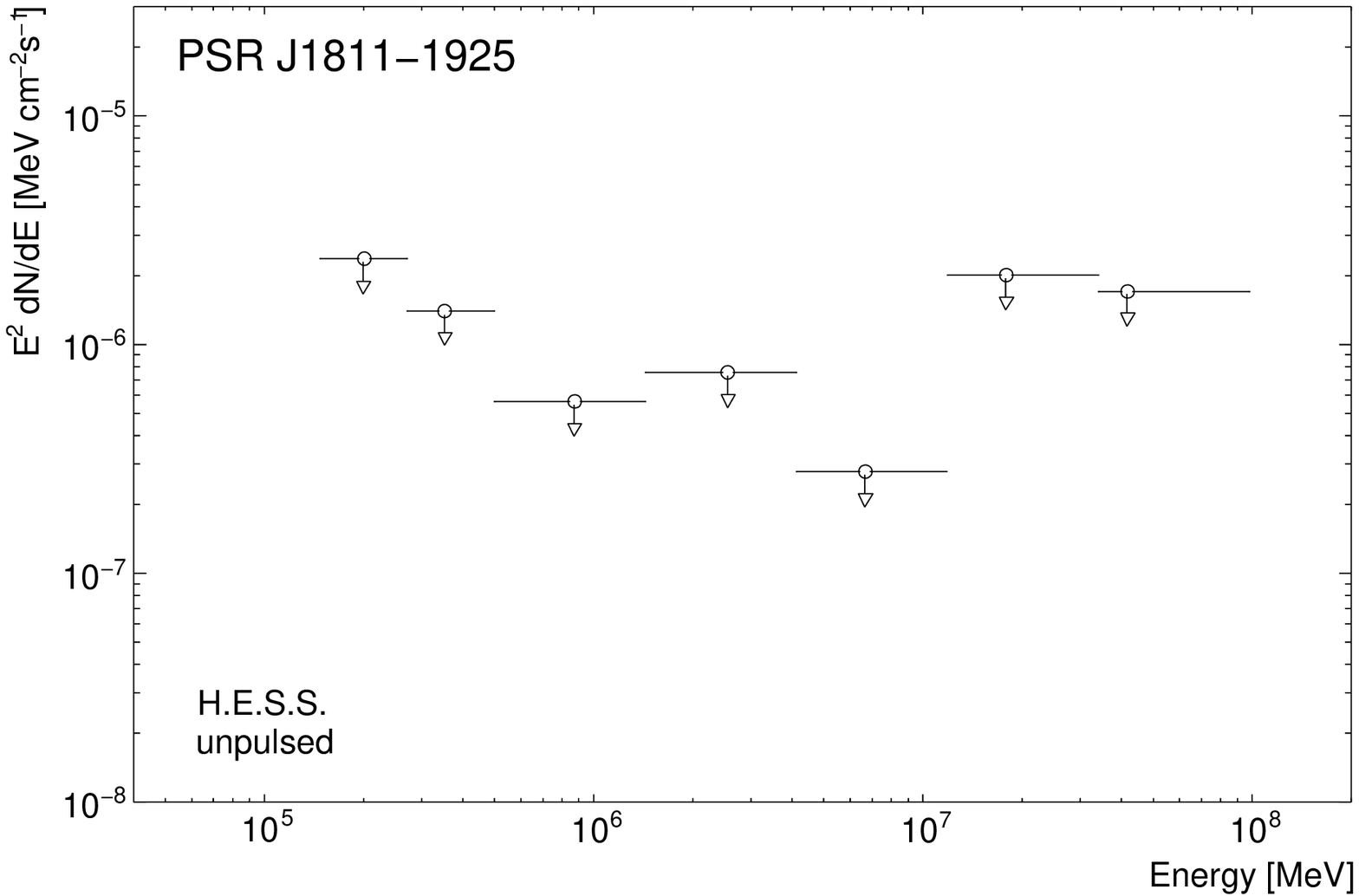}}

\caption{
Differential upper limits at 99\% confidence on the energy flux of
unpulsed VHE \g-ray emission for PSR\,J1747$-$2958, PSR\,J1801$-$2451
and PSR\,J1811$-$1925 (from top to bottom) as measured by H.E.S.S.}
\label{fig_spectra_unpulsed}
\end{figure}

Differential energy flux upper limits (i.e.\ in $E^2{\rm d}N/{\rm
d}E$) were derived for each data set, for the energy range from
threshold energies up to the highest energies for which significant
event statistics were accumulated. In
Figures~\ref{fig_Crab_spectrum}--\ref{fig_spectra_unpulsed} the
differential flux limits calculated are shown separately for each
pulsar. Different points are shown for the different limit
determination methods (see Sec.~\ref{sec_limits}). The points above
energies of 0.5\,TeV correspond to the results for the standard
selection cuts and the points below 0.5\,TeV to these for the low
energy cuts. The horizontal position of the points denotes the mean
energy of the events in each energy bin while the bin edges are
indicated by the horizontal error bars. In the case of the pulsed
fraction method, the mean energy was calculated for the assumed power
law energy spectrum.

Note that the overall \g-ray energy scale for the standard cuts has a
systematic error estimated to be 20\% for a power law energy spectrum
with $\Gamma$\,$\sim$\,$2.6$ \citep{Crab}, dominated by the absolute
shower energy calibration of the instrument. The impact of the
systematic uncertainty in the energy reconstruction method on the flux
upper limit for the low-energy cuts was determined to be 30\% for
energies around the energy threshold of the respective data set,
mainly affecting the given flux limit point for the lowest energy bin.

\section{Discussion}
\label{sec_discussion}

For the four \g-ray pulsars, these new results allow some general
conclusions on high-energy emission mechanisms of pulsars to be
drawn. The results are put in the context of model predictions from
selected polar cap and outer gap scenarios. In the particular case of
the outer gap scenario, only those model calculations were considered
which make explicit predictions for the IC component at TeV
energies. Where appropriate,
simulated spectra are shown which are calculated using simulation
codes from \citet{pc:3} and \citet{og:2}. Finally, the results can be
compared with findings of other experiments in the TeV energy domain.

\begin{figure}[t]
\centering
\resizebox{\hsize}{!}{\includegraphics{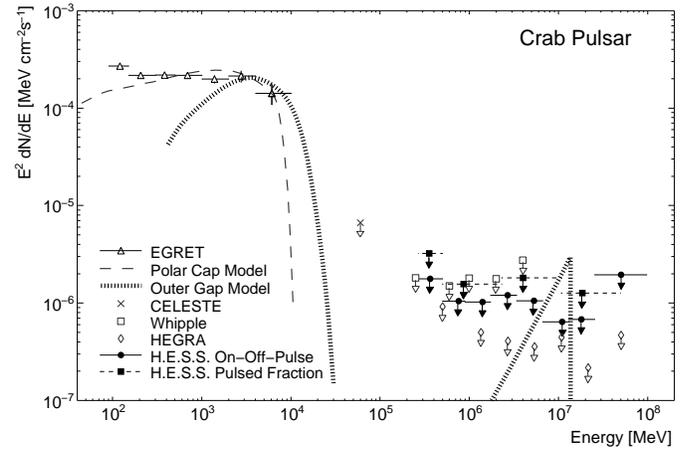}}
\caption{
H.E.S.S.\ energy flux limits (99\% c.l.) for pulsed emission of the
Crab pulsar. The full circles and full squares correspond to the
on-off-pulse and pulsed fraction limit determination methods,
respectively. Below energies of 0.5\,TeV the results were obtained
with the low energy selection cuts, otherwise the standard cuts were
used. Note that the threshold energy of the H.E.S.S.\ data lies at
310\,GeV due to the observation at high zenith angles imposed by the
limited accessibility of the northern sky. The crosses, open squares,
and open diamonds denote limits from CELESTE \citep[99\%
c.l.,][]{celeste_crab}, Whipple \citep[99.9\% c.l.,][]{whipple_crab},
and HEGRA \citep[99.865\% c.l.,][]{hegra_crab:2}, respectively. The
open triangles show the EGRET phase-averaged spectrum
\citep{egret_crabvela}. The indicated polar cap curve was generated
according to \citet{pc:3} and the outer gap model curve taken from
\citet{og_crab}.}
\label{fig_Crab_spectrum}
\end{figure}

\begin{figure*}[ht]
\centering
\resizebox{\hsize}{!}{\includegraphics{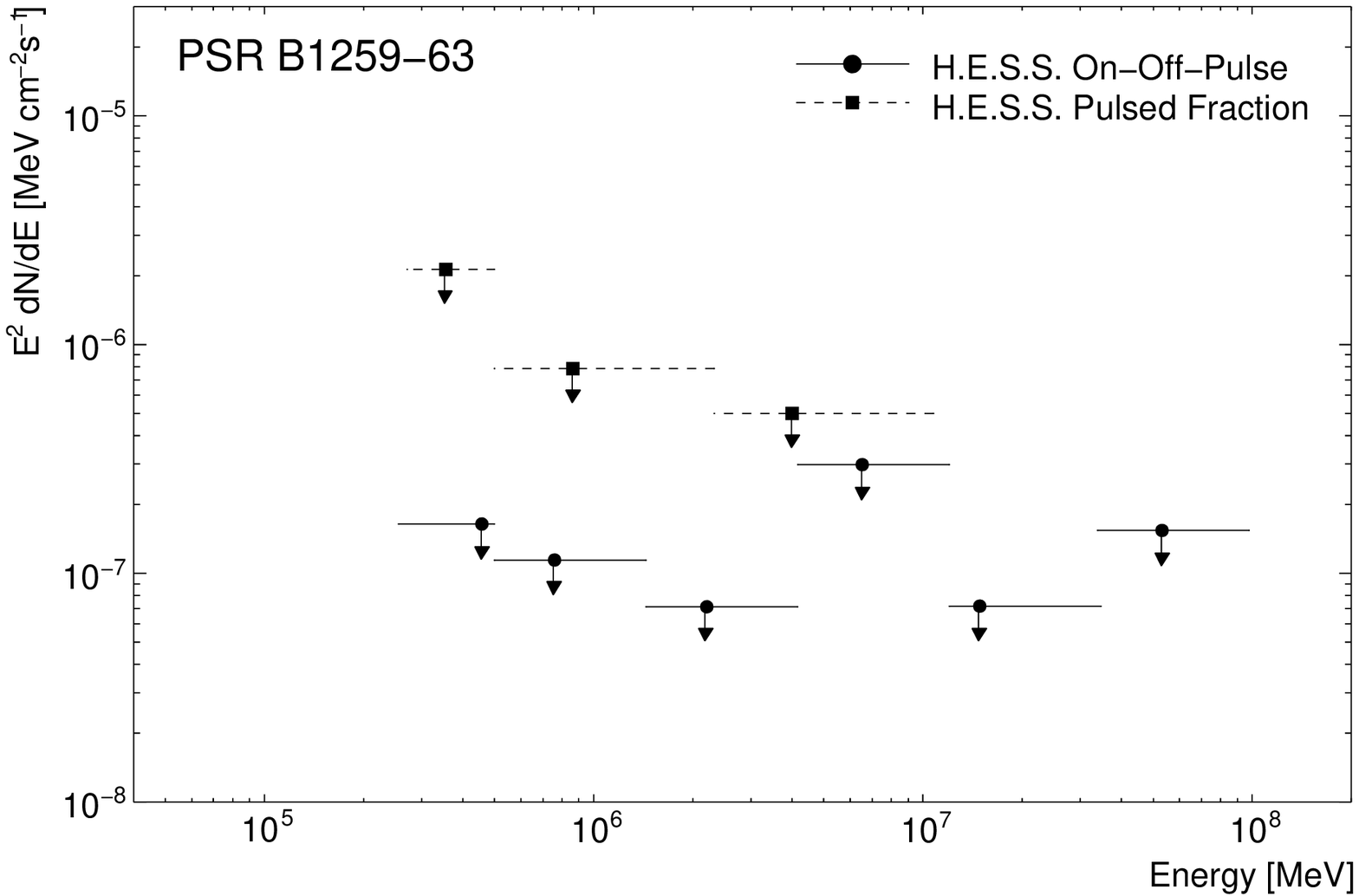}
\includegraphics{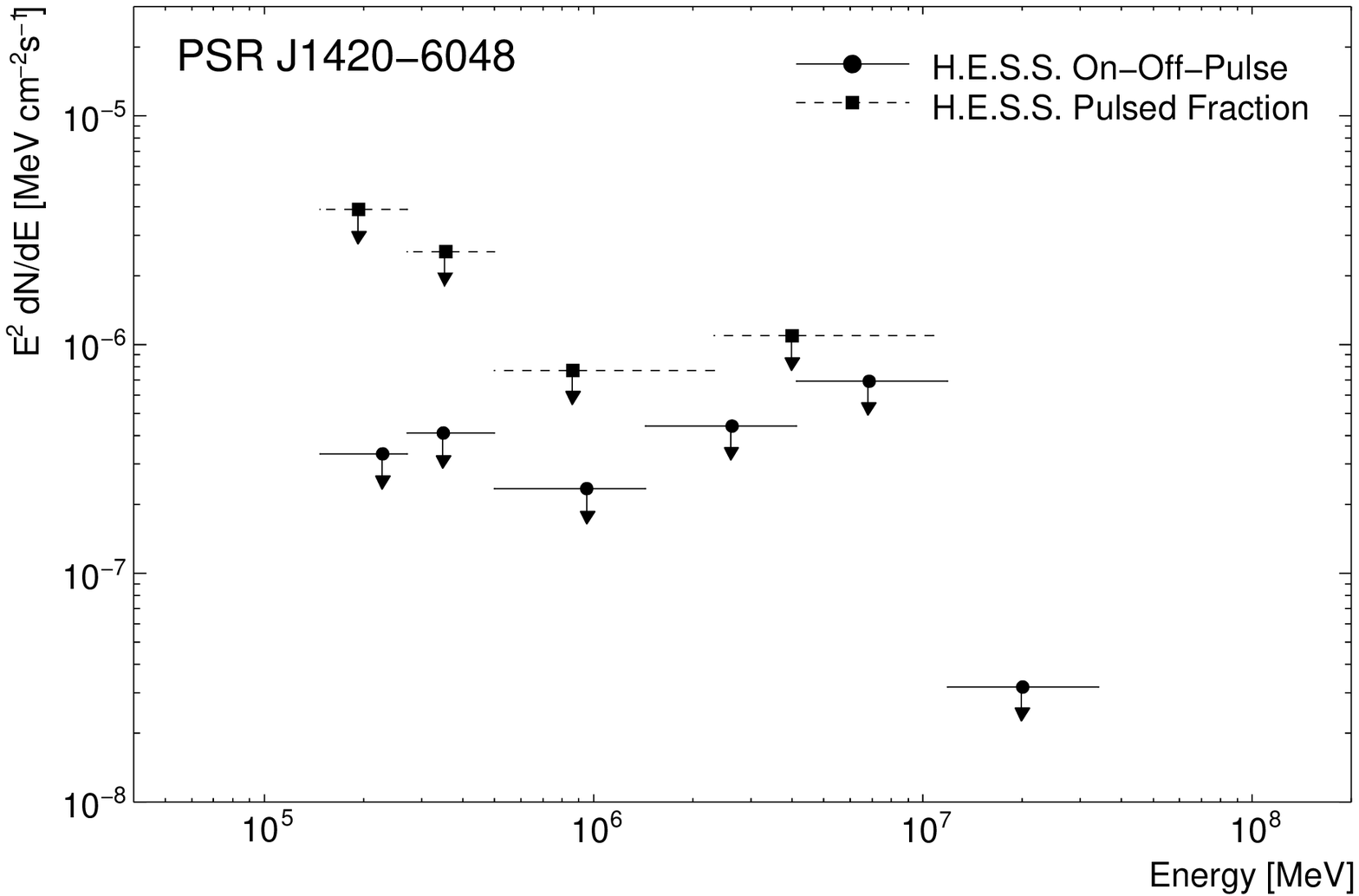}}
\resizebox{\hsize}{!}{\includegraphics{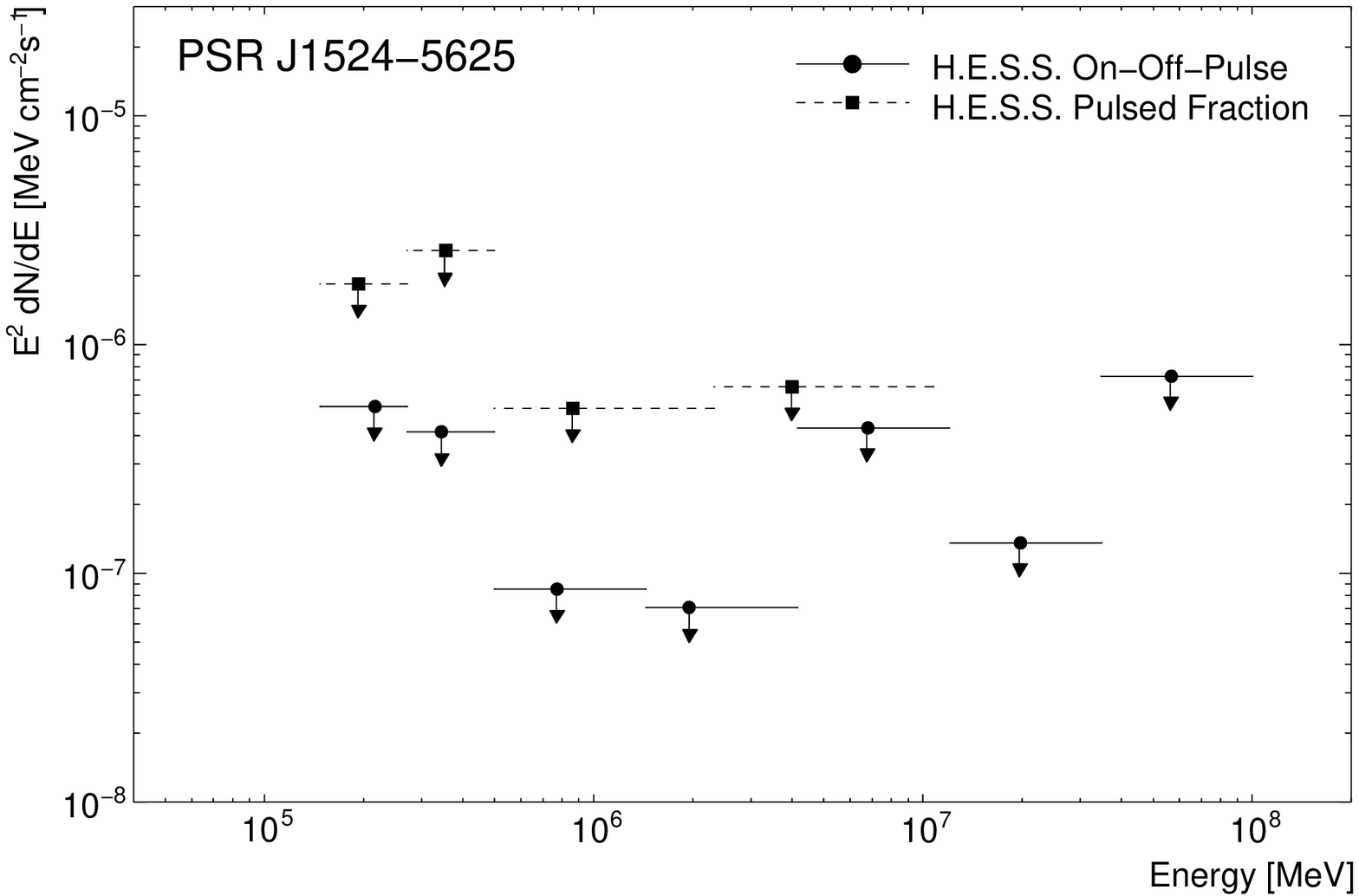}
\includegraphics{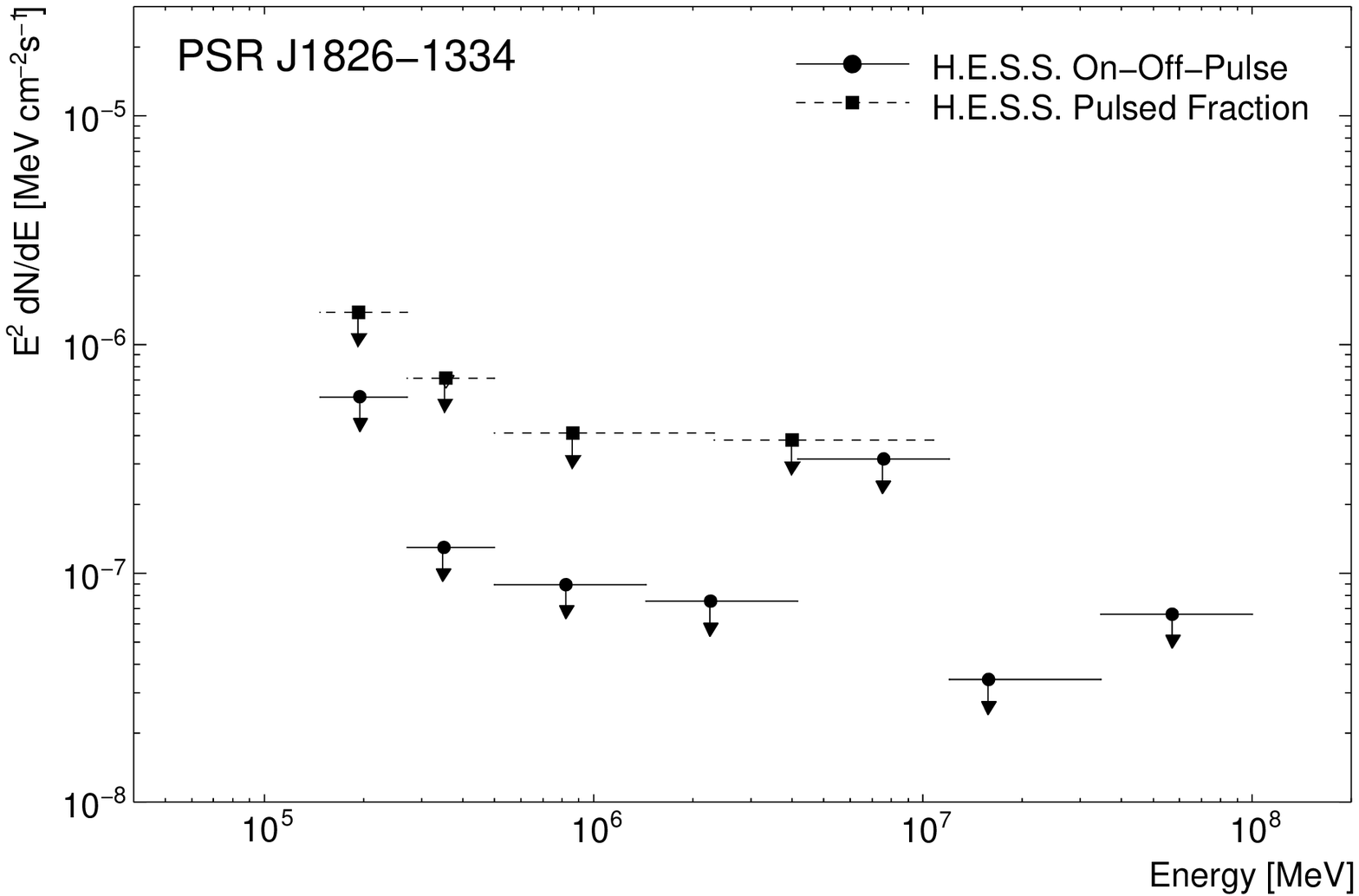}}

\caption{
Differential upper limits on the energy flux of pulsed VHE \g-ray
emission for non-\g-ray pulsars as measured by H.E.S.S.\ (at 99\%
confidence, point descriptions as in Fig.~\ref{fig_Crab_spectrum}).}
\label{fig_spectra}
\end{figure*}

{\bf Crab pulsar}. The Crab pulsar has been studied by all major
ground-based \g-ray instruments at energies between 60\,GeV
\citep{celeste_crab} and 50\,TeV
\citep{hegra_crab:2}. Figure~\ref{fig_Crab_spectrum} shows the new
H.E.S.S.\ results along with \g-ray measurements from EGRET and
selected results in the VHE domain. The southern location of H.E.S.S.\
allows only observations at rather high zenith angles, prohibiting a
deep exposure especially at low energy thresholds. Therefore the
limits obtained using the Northern Hemisphere instruments HEGRA,
with more than 200\,hours of observations, and Whipple, with a
lower energy threshold \citep{whipple_crab}, could not be improved
upon. However, the HEGRA limits are derived only for a rather
narrow pulse region between $-0.06$ and 0.04 of the main pulse,
whereas the Whipple and H.E.S.S.\ limits include both the main
and secondary pulse regions. Taking the H.E.S.S.\ results for the
on-off-pulse method and assuming a power-law spectrum with exponential
cutoff ${\rm d}N/{\rm d}E\propto E^{-\Gamma}\exp(-E/E_{\rm c})$, with
$\Gamma=2.05$, the cutoff energy is constrained to be $E_{\rm
c}$\,$<$\,$80$\,GeV, less constraining than the flux limit at
$\sim$$60\,$GeV reported by CELESTE \citep{celeste_crab}. Our
new differential flux upper limits for the Crab pulsar constrain the
IC component in the TeV range predicted by \citep{og_crab},
reinforcing the constraints placed by previous \g-ray experiments,
both at GeV and TeV energies. No constraints can be placed on the
polar cap prediction, which exhibits a sharp cutoff at a few GeV.

\begin{figure}[t]
\centering
\resizebox{\hsize}{!}{\includegraphics{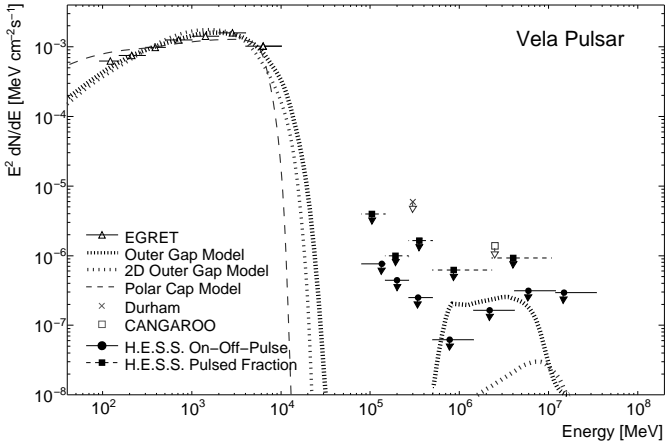}}
\caption{
H.E.S.S.\ energy flux limits (99\% c.l.) for pulsed emission of the
Vela pulsar (see Fig.~\ref{fig_Crab_spectrum} for point
descriptions). The cross and open square denote limits from the Durham
Mark 6 telescope \citep[99.865\% c.l.,][]{durham}, and the CANGAROO
detector \citep[95\% c.l.,][]{cangaroo_vela}, respectively. The open
triangles resemble the EGRET phase-averaged spectrum
\citep{egret_crabvela}. The polar cap and outer gap model curves are
generated according to the model of \citet[][ black dotted
curve]{pc:3} and from \citet[][ dashed curve]{og:2}, respectively. The
dotted grey outer gap model curve is taken from \citet{og2d}.}
\label{fig_Vela_spectrum}
\end{figure}

{\bf Vela pulsar}. For Vela, the limits set using the Durham Mark 6
telescope \citep{durham} and the CANGAROO \citep{cangaroo_vela}
detector are improved by more than one order of magnitude in flux and
reach down to $\sim$$100$\,GeV (see Fig.~\ref{fig_Vela_spectrum}),
constraining the cutoff energy for the EGRET power law with
$\Gamma=2.38$ to be $E_{\rm c}<14$\,GeV. Additionally, the H.E.S.S.\
limits significantly constrain the parameter space for the IC
component of the outer gap model \citep{og:2}, which was computed with
a slight change of the gap size $D_\perp=0.177$ \citep[see][ for
details]{og:2} to better match the EGRET data. Note that in the
model calculation of \citet{og:2}, no contribution from the photons of
the optical pulse to the outer gap seed photons was considered
\citep{hirotani_okkie} which would result in a further increased IC
component. On the other hand, a more detailed treatment of the outer
gap anisotropy would lead to a decreased IC contribution
\citep{hirotani_okkie}. A newer, two-dimensional version of the same
outer gap model, which includes the contribution from the optical and
infrared pulse, predicts a lower IC flux in the TeV range
\citep{og2d}. The polar cap model again predicts pulsed emission
only at energies far below the energy threshold of current ground-based
\g-ray detectors.

\begin{figure}[t]
\centering
\resizebox{\hsize}{!}{\includegraphics{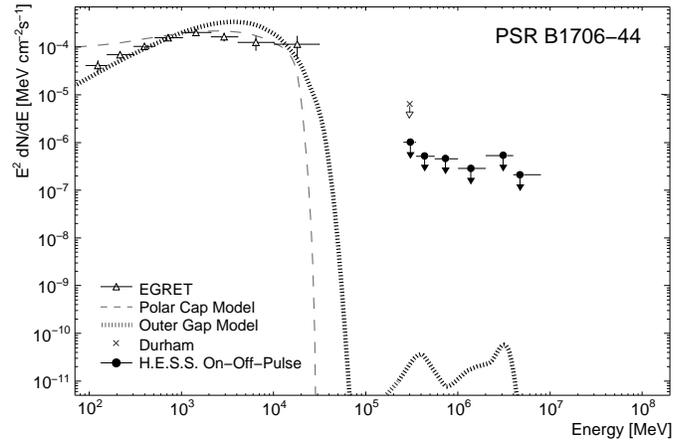}}
\caption{
H.E.S.S.\ energy flux limits (99\% c.l.) for pulsed emission of
PSR\,B1706$-$44 for the on-off-pulse method (full points). The cross
denotes a flux limit from the Durham group \citep[99.865\%
c.l.,][]{durham}. The open triangles show the EGRET phase-averaged
\g-ray spectrum \citep{egret_1706}. Polar cap and outer gap model
curves were calculated according to \citet{pc:3} and taken from
\citet{og:2}.}
\label{fig_1706_spectrum}
\end{figure}

{\bf PSR\,B1706$-$44}. For this pulsar, limits on pulsed \g-ray
emission were previously reported by \citet{durham}. Although
H.E.S.S.\ data were mostly limited to 2-telescope data from the
commissioning phase to which only the on-off-pulse method has been
applied, more restrictive limits could be set (see
Fig.~\ref{fig_1706_spectrum}), constraining the exponential cutoff
energy to be $E_{\rm c}<71$\,GeV using the power law fit of the EGRET
spectrum with $\Gamma=2.25$. However, the IC component in the outer
gap scenario of \citet{og:2} is left unconstrained by several orders
of magnitude.

\begin{figure}[t]
\centering
\resizebox{\hsize}{!}{\includegraphics{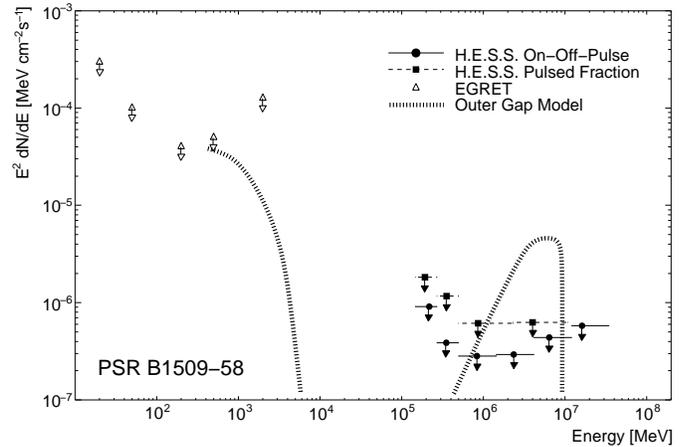}}
\caption{
H.E.S.S.\ energy flux limits (99\% c.l.) for pulsed emission of
PSR\,B1509$-$58 (see Fig.~\ref{fig_Crab_spectrum} for point
descriptions). The open triangles resemble the EGRET upper limits for
pulsed emission \citep{egret_limits}. The outer gap model curve
(dotted) originates from \citet{og_1509}, Fig.~6.}
\label{fig_1509_spectrum}
\end{figure}

{\bf PSR\,B1509$-$58}. The rather strong unpulsed VHE \g-ray
emission of the PWN surrounding PSR\,B1509$-$58 decreases
the flux sensitivity for pulsed emission in comparison with other
pulsars. The limits shown in Fig.~\ref{fig_1509_spectrum} represent
the first sensitive search for sub-TeV \g-rays from this
pulsar. The IC component of an outer gap model prediction
\citep[][ dotted curve, parameters $j_1=0.6, j_2=0$]{og_1509} can be
ruled out by the H.E.S.S.\ limits, although the curvature radiation
component seems to be already marginally constrained by the EGRET
limits at energies $<200$\,MeV.

The remaining pulsars have not been detected in the MeV--GeV range and
no detailed calculations based on any high-energy emission models are
available for the TeV range.

Independent of any \g-ray emission scenario, for all selected pulsars
the flux limits obtained imply an upper limit on the ratio of the
total pulsed VHE \g-ray luminosity compared to the pulsars' spin-down
power in the range $10^{-6}$ to $10^{-4}$. This limit on the \g-ray efficiency is significantly lower
than the corresponding EGRET sensitivity in the GeV range.

\section{Conclusions}


H.E.S.S.\ data on young and energetic pulsars, including four of the
eight known GeV \g-ray pulsars, have been searched for evidence of
pulsed \g-ray emission at energies between 100\,GeV and 100\,TeV. No
conclusive evidence for pulsed emission has been found and
differential upper limits on the pulsed flux were derived using a
variety of different methods, constraining the pulsed flux for a wide
range of possible pulse shapes and spectra in the VHE \g-ray range.

In the case of the well-studied \g-ray pulsars Crab, Vela, and
PSR\,B1509$-$58, the pulsed flux upper limits reported here
significantly constrain the inverse Compton component of selected
outer gap models for flux predictions in the TeV range which were
available at the time of writing. However, the results cannot rule out
the outer gap scenario in general, since not all of the different
calculations published give predictions for the IC component.
Also, free model parameters such as seed photon density,
inclination, gap geometry or viewing angles can be adjusted for the
model cases considered here. However, it remains an open challenge for
theory to provide reasonable parameters consistent with the flux
limits reported here.

For the complete sample of pulsars, containing a large fraction of the
most energetic pulsars currently known, the absence of pulsed VHE
\g-ray emission already disfavours a significant contribution of the
IC component to the energy loss mechanism of these pulsars. This
sample exhibits great variety in different pulsar characteristics,
providing valuable input to further studies of the outer gap IC
component. For example, in the case of the binary pulsar
PSR\,B1259$-$63, a large spin-down luminosity is accompanied by a
rather weak magnetic field, i.e.\ reduced \g-ray absorption by pair
creation, and additionally the photon field originating from the
companion star provides a high density target with known energy
spectrum for a possible pulsed IC component, fixing an otherwise
uncertain model parameter.

Although in several cases there is spatial coincidence with extended
TeV \g-ray emission, pulsed emission is not detected in VHE \g-rays.
In particular, the flux upper limits derived are of the order of
$10^{-4}$ to $10^{-6}$ of the pulsar spin-down flux, underlining the
non-magnetospheric origin of the TeV radiation component and
supporting the widely accepted scenario of an effective energy
transport mechanism to, and strong particle acceleration in, the
pulsar wind nebula.

Finally, the upper limits given here imply a steep turnover of the
pulsed high-energy spectrum at energies of a few tens of GeV for all
pulsars studied which have established high-energy ($>$\,keV)
emission, confirming a prediction common to all pulsar
models. Moreover, for the remaining pulsars, a turnover at sub-TeV
energies would be likely even if pulsed GeV \g-ray emission with
spectral properties similar to that of the established \g-ray pulsars
is detected with future sensitive instruments
\citep{GEVprospects}. However, the different scenarios for pulsar
emission differ significantly in their predictions of the exact shape
and energy of the turnover. Thus, the search for pulsed \g-ray
emission from pulsars provides interesting prospects for future
satellite-based and low-threshold ground-based \g-ray instruments.


\begin{acknowledgements}

The support of the Namibian authorities and of the University of
Namibia in facilitating the construction and operation of H.E.S.S.\ is
gratefully acknowledged, as is the support by the German Ministry for
Education and Research (BMBF), the Max Planck Society, the French
Ministry for Research, the CNRS-IN2P3 and the Astroparticle
Interdisciplinary Programme of the CNRS, the U.K. Particle Physics and
Astronomy Research Council (PPARC), the IPNP of the Charles
University, the South African Department of Science and Technology and
National Research Foundation, and by the University of Namibia. We
appreciate the excellent work of the technical support staff in
Berlin, Durham, Hamburg, Heidelberg, Palaiseau, Paris, Saclay, and in
Namibia in the construction and operation of the equipment.

\end{acknowledgements}

\bibliographystyle{aa}
\bibliography{pulsed}

\end{document}